# Keyword Search Engine Enriched by Expert System Features

## Olegs Verhodubs

oleg.verhodub@inbox.lv

**Abstract**. Keyword search engines are essential elements of large information spaces. The largest information space is the Web, and keyword search engines play crucial role there. The advent of keyword search engines has provided a quantum leap in the development of the Web. Since then, the Web has continued to evolve, and keyword search systems have proven inadequate. A new quantum leap in the development of keyword search engines is needed. This quantum leap can be provided with more intellectual keyword search engines. The increased intelligence of such keyword search engines can be achieved through a combination of keyword search engines and expert systems. The paper reveals how it can be done.

## 1      INTRODUCTION

Over the last few centuries, technological progress has been the driver of the development of our civilization. All processes had an informational dimension and technological progress was not an exception. It consisted in an ever-increasing release of information into the information space. Initially, the information space was a multitude of newspapers, books, lectures at universities. Then the Web appeared and the information space expanded significantly. The amount of information in the Web has been constantly increasing, and this has led to difficulties in finding the necessary information. Keyword search engines came to the rescue. Keyword search engine is a computer program that looks for keyword match in its database and displays the documents, where the match is greatest. Certainly, the database has to be filled with information before using keyword search engines. In general, it is possible to talk about different keyword search engines, but in the paper, keyword or keyword-based web search engines are kept in mind, whenever keyword search engines are mentioned in the paper. So, keyword-based web search engine is a keyword search engine that operates with data in the Web. There are many keyword-based web search engines, and almost everyone uses them every day. They are GOOGLE [1], BING [2], BAIDU [3] and others. Despite the large number of keyword-based web search engines and some of their differences, all of keyword-based web search engines work the same way. The user types the keywords in the appropriate field, then the user presses the button and after that, the keyword-based web search engine displays the list of links to documents in the Web, where the typed keywords are most often found. This was enough for a long time. Nevertheless, the amount of information in the Web has been constantly increasing, and the list of links produced by keyword search engine has been getting longer. The longer the list of links is, the much time is necessary to view and analyze it. It is unacceptable to be putting the information together and draw conclusions based on it in the absence of time. The capabilities of keyword search engines have become insufficient. It is possible to increase the efficiency of keyword search engines by moving in several directions. The first direction is the development of methods for putting the information together from various sources so that as a result of a search by keywords a logically verified text is generated, composed of facts obtained from different information resources. In this case, it would not be necessary to read all the documents from the found links. The second direction aims at deeper intellectual processing of information in the Web. This can be done in many ways. One way is extracting knowledge from the Web and then inferencing based on this knowledge. As a result, new facts are being available. Knowledge, inferencing, new facts are features specific to expert systems, so we speak about keyword search engines enriched by expert system features.

This paper composed of several sections. The next section describes the information space of keyword search engines. The third section sets out the possibility of keyword search engine enrichment with expert system features. This allows to expand the result of work of keyword search engine by getting more information, than is indexed in the usual keyword search engine database. The following section presents the way of rule compressing in the keyword search engines with expert system features. Actually, being presented rule compression is useful for many rule-based systems.

## 2      INFORMATION SPACE

Information space is an information environment, where systems exist. Systems are different and all of them are surrounded by information, but only computer systems are considered here. Computer systems and also other systems use and produce information. Keyword search engines exist in their own information space. The information space of keyword search engines consists of the user request, database of indexed information and the result. The result is the result of keyword search engine, namely information that is produced by keyword search engine based on the user request and indexed information in the database. Information saved in the database is being obtained by indexing documents in the Web. Indexing documents means saving information about documents in the Web and keywords that are found in these documents. The user request is a set of keywords that user enters into the keyword search engine to search for matches of keywords and indexed information in the database. If entered keywords match the indexed information, the system outputs the list of documents, where these keywords are found. Typically, the list is ranked by keyword match frequency. The principles of output ranking are interesting in terms of science, but the structure of user request is no less interesting scope that is why this scope is discussed in details further.

The structure of the user request is complicated, but in the simplest case, this structure is the following:

$$Q \in \{q_1, q_2, q_3, ..., q_n\}, \tag{1}$$

where $Q$ is a user request and $q_1,\ q_2,\ q_3,\ ...,\ q_n$ are components of the user request, namely keywords. In this case, all keywords of the user request are tantamount. This means that the keywords can be placed in any order and different order of keywords will not affect the result of the keyword search engine work. This is an ideal case, which almost never happens in real life. In real life, the keywords of the user request are not tantamount and the structure of the user request takes the following form:

$$Q \in \{q \mid P(q)\}, \tag{2}$$

where $Q$ is a user request, $q$ is a set of the keywords in the user query, $P(q)$ is a set of keyword attributes. The set of keyword attributes defines the structure of the user request. This structure allows forming the result of the work of the keyword search engine more precisely. The more attributes are; the more levers are to filter data for result.

There are a lot of attributes that can be associated with keywords in user request. It is natural for humans to highlight the main thing and push into the background the secondary, thus it is about different priorities. In theory, the user can input the value of priority together with each keyword in a user request, but in practice, it is much more convenient to use the order of keywords in the user request for detecting the priorities of keywords. Usually, the priority of keywords in user request decreases from left to right, but the order may be reversed that is from right to left.

Another attribute that can be associated with keywords in user request is a search history. This attribute implies a semantic connection between consecutive user requests. For example, if a user looks for information about movies released in 1991 in its first request and after that looks for information about the actor Schwarzenegger in its second request, then there is reason to believe that there is a semantic connection between these requests and the desired result is most likely related to the "Terminator 2" movie. A semantic connection between user requests is something tangible and generally easy to calculate. An emotional connection of consecutive user requests is not so tangible and difficult to distinguish or calculate. One of the reasons for this is the difficulty in distinguishing between a temporary emotional state and a psychological type of user identity.

The history of user requests and the priority of keywords form the structure of the user request, but this structure is homogeneous that is all keywords belongs to one semantic group. Such user requests are being used, but they are not all possible user requests. Another type of user requests is complicated requests with heterogeneous structure that is such user requests, where keywords belong to different semantic groups. For example, the user request, where its keyword set includes not only objects, but also the properties of these objects. In such a request the objects and their properties belongs to different semantic groups; the properties are subordinate to the objects. The objects and their properties is not the only relation among keywords in the user request. For example, user request can consist of keywords, which defines the objects, only, but some of these objects can relate to other objects by means of "part_of" relation. Alternatively, user request can consist of keywords, which defines the objects, but some of these objects can relate to other objects by means of "is_a" relation. In such a case, the complicated user request is a fact or information that should be confirmed, or disproved by resources in the Web. The difference between facts and information is that information is a set of semantically interrelated facts [4].

## 3    EXPERT SYSTEM FEATURES

One more type of user requests is a request, where its keywords are attributes of some not mentioned object. For example, pointing "wings", "engine" and "landing gear" in the keyword search engine, a plane is implied. In this case, the user is not interested in the list of web documents, where "wings", "engine" and "landing gear" words are met; highly likely, the user would like to identify the object with the listed attributes.

Another type of user requests is a request, where its keywords are some objects with a possible logical connection. For example, listing the objects "passenger car" and "vehicle" in the request, the user hopes to establish a logical connection between the listed objects that is the user intends to find out if "passenger car" is a "vehicle".

Or if a user types "fish" and "lives in", he would prefer to get the "water", or "lake", or "sea" as a result, but not the list of the web documents, where "fish" and "lives in" are met most often.

Of course, there are a lot of other types of complex requests, where objects, attributes, relations may be present and the user is not interested in the list of web documents, where the mentioned keywords are present, but is interested in some intellectual superposition of mentioned keywords. Here intellectual superposition is something semantically common for mentioned keywords. The clearest way to reach this is to exploit the area of artificial intelligence namely the area of expert systems. The area of expert systems explores the ways to develop expert systems. An expert system is a computer system that emulates the decision-making ability of a human expert [5]. There are many different classifications of expert systems. Expert systems can be paid or free, autonomous or integrated, static or dynamic. From the developer point of view, expert systems are systems that are based on one of the knowledge representation model. To be more precise, a knowledge base is built on some knowledge representation model, where the knowledge base is a part of an expert system. There are several knowledge representation models such as formal logical model, frames, semantic networks and production model [6]. A production model also called rule-based model, because knowledge in this model is represented in the form of rules. Rules are sentences in the form of "IF condition, THEN conclusion". A rule acts as follows: if condition is true, then conclusion is true. Condition and conclusion are some statements or facts. It is possible to inference using available facts and rules. The process of inferencing means obtaining new facts using rules and existing statements. In general, there are two types of inferencing: forward chaining and backward chaining [6]. Forward chaining starts with the available facts and rules to obtain new facts. Forward chaining goes on until a goal is reached. On the contrary, backward chaining starts with the goal or a list of goals and goes in the opposite direction. Despite the differences in forward and backward chaining, the key point is that facts and rules are able to provide us with new facts. This is very useful for improving the performance of keyword search engines. In addition to concentrating the collected information, producing new facts is the most natural way to improve keyword search engines. Here new or produced facts have less confidence than facts composing the collected information. This has to be clear to the user of keyword search engine that is why it is necessary to differ the output information namely dividing this information into at least two categories. These categories are reliable facts and produced facts. An example of the desired output of the keyword search engines might look like this (Fig.1):

**Fig.1.** Result of keyword search engine work.

Here a user types the keyword "napoleon" in the field of the keyword search engine and then presses the "start" button (Fig. 1). In response, keyword search engine provides information divided into two categories: facts and conclusions. Facts are something that is found in the Web. Conclusions are statements that obtained based on the facts and rules. The rules that are applied for the listed facts may be the following:

IF *<E1b1b1 haplogroup>* THEN *<ancestors from the Middle East>*

IF *<conquer>* THEN *<unifier>*

IF *<arsenic in hair>* THEN *<was poisoned>*

One more example is shown in Fig.2.

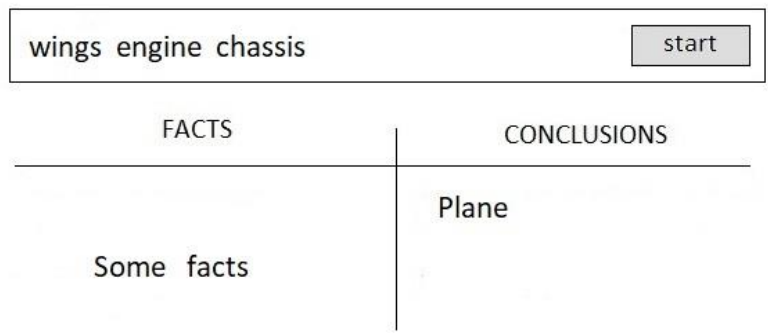

**Fig.2.** Result of keyword search engine work.

A user types "wings engine chassis" keywords and then presses the "start" button. Keyword search engine outputs some facts and the conclusion, which is based on the following rule:

IF *<wings* and *engine* and *chassis>* THEN *<plane>*

In general, speaking about expert systems there are two basic operation modes of them: the knowledge acquisition and the consultation modes. The knowledge acquisition mode aims to fill knowledge of the expert system. The consultation mode aims to get a consultation from the expert system. Sometimes one more operation mode of expert systems is distinguished. This is an explanation mode [7]. This mode allows the expert system to explain its conclusions and its reasoning process. Nevertheless, explanation of reasoning process is present in each expert system and even if this mode is not provided explicitly, it is present as a part of the reasoning process itself. That is, the process of reasoning is accompanied by the information, how this reasoning has been done. Unlike expert systems, where consultation with the expert system occurs in the form of a question and answer, basically keyword search engine enriched by expert system features do not need this form of interaction with a user. All needed information has already been inputted in the field for keywords. Only the rules directly applicable to the information entered by the user are utilized. These rules have been generated before for area, which is detected based on the entered keywords. Keyword search engine enriched by expert system features can precise some information from the user, but it is supposed that this happens not often.

Regarding the implementation, such a keyword search engine with expert system features could be realized in several ways. The easiest way to implement this is to exploit the technologies of the Semantic Web. Here OWL (Web Ontology Language) is a technology of the Semantic Web that can help. The OWL language is intended to describe ontologies [8]. Ontologies are used to specify some of domains as medicine, information technology, construction, economics, or something else. It is possible to specify websites in the form of OWL ontologies, too. Classes, relationships between classes, attributes and class instances (individuals) are main elements that are available in terms of the OWL language to specify some of domain. This is enough to realize keyword search engine enriched by expert system features. Classes and instances of these classes in the ontology are used to output facts, but rules are responsible for inferring conclusions. Rules are generated from OWL ontologies as presented in [9, 10]. The bottleneck of this way is an ontology generation from the website. The task of ontology generation from the website is related to the task of ontology generation from text and is called ontology learning. Many ongoing researches in this area indicate that the quality of generated ontology is not excellent yet. One more way to implement keyword search engine with expert system features implies knowledge generation from the text. In this case, OWL ontologies are not critical for the keyword search engine enriched by expert system features. The task of knowledge generation from text is developing now, and it is not presented here.

## 4      RULE COMPRESSION

Rules are useful for enhancing the keyword search engines; because by means of rules the output of keyword search engines become more intelligent. Nevertheless, use of rules has some complexities that have to be resolved. Guidelines are necessary to know how to resolve these complexities. Guidelines are in other words meta-rules, since meta-rule   is   a   rule that describes how other rules should be used   [11]. The   first complexity is a choice of conflict resolution strategy. A conflict resolution strategy is required to make the decision as to which rule should be fired first [12]. There are several strategies here. One of the strategies is rule ordering that is first rule is served the first. One more strategy implies firing the rule with most conditions attached. There are other strategies, however it is not necessary to overview all of them if existing inference engine is being used. Inference engine is a program that models expert's style of reasoning using knowledge from the knowledge base [13]. To be more precise, inference engine is not always a program, but also a software library that is easy to connect and to use. A software library for use with the semantic technologies is Apache Jena [14]. It has several inference engines as RDFS reasoner, OWL reasoner, transitive reasoner and general purpose rule engine. The general purpose rule engine is convenient, when there are different sources for rule generation. For example, these sources are ontologies and raw text in websites. The general purpose rule engine of Apache Jena can be also adjusted for fuzzy reasoning as was shown in [15]. This gives an opportunity to cope with the lack of credibility in the Web.

The next complexity using rules is a length of reasoning. In expert systems reasoning has been continuing while facts and rules are available, but this is not always necessary for keyword search engine with expert system features. It is necessary to enable only those rule sequence that helps to  respond a user request most accurately. Even if there are more facts and rules, it is not necessary to enable these rules, if the conclusions of them do not reveal something new related to the request of the user. That is the number of enabled rules is limited to the semantic area of the conclusions of these rules. For example, suppose a user enters the keywords "car logistics" and there are the following rules are available:

IF *<car>* THEN *<used for transportation>*

IF *<car>* THEN *<used in race>*

IF *<used for transportation>* THEN *<logistics>*

IF *<used in race>* THEN *<car is special>*

Only the first and the third rules are interesting to the user based on the set of entered keywords, because the conclusions of these rules are semantically correlates with the entered keywords. In real life, the set of available rules can consist of more rules, and therefore the task of rule filtering is much more important.

In general, each sequence of keywords enables a certain number of rules. This set of rules produce the set of certain conclusions. Therefore, it is not necessary to reason every time the user enters a certain set of keywords if rules for these keywords do not change. It is possible to reduce the sequence of rules to one rule only as is shown below:

IF <*a*> THEN <*b*>

IF <*b*> THEN <*c*>

IF <*c*> THEN <*d*>

IF <*d*> THEN <*e*>

Here the sequence of 4 rules is reduced to one single rule:

IF <a> THEN <e>

However, this is not the only way to reduce these four rules to one rule. For example, such an option has the right to exist, too:

IF <*a*> THEN <*d*>

It is clear that there are other ways of reducing several rules to single rule. The way is being selected based on the area of the typed keywords. The more the conclusion of the rule matches the area, the more reason to enable the rule. On the other hand, it may happen that the conclusions of all rules satisfy the area of the typed keywords, but in this case, we of course cannot speak about reducing of rules.

The possibility of rule count reducing is technological in the sense that it provides the higher productivity. The higher productivity means here that there is no need to reason from start to finish every time a user types a certain sequence of keywords. It would be enough if every sequence of keywords would have an associated set of rules. Technically, this can be done using a database, where sequences of keywords are stored with the associated set of rules.

## 5      CONCLUSION

The development of keyword search engines has run up against a wall of new challenges born of the need for increased intelligence in search engines. Only increased intelligence can let the keyword search engines to stay useful, what means the higher quality of result filtering in keyword search engines. As you know, keyword search engines output the list of addresses, where information about inputted keywords is situated. Increased intelligence or the higher quality of result filtering in keyword search engines means that information about the inputted keywords is accumulated from everywhere in the Web into the whole. Besides, a keyword search engine with the increased intelligence provides inferencing based on the available keywords, data and rules. Data, rules and the possibility of inferencing are attributes of expert systems namely rule-based expert systems. Thus, such a keyword search engine with the increased intelligence is named as keyword search engine enriched by expert system features. Rules are key part of rule-based expert system and there are several ways how rules can be obtained. The most explored way for rule generation is using OWL ontologies for this purpose, however there are other ways for rule generation, too. OWL ontologies are still not widespread that is why other ways of rule generation can be useful. On the other hand, there is no need to inference from the beginning to the end always, using all available data and rules. It is much more often required to use one-two-three rules and some data in order to make the simplest conclusions. That is why we speak about keyword search engine enriched by expert system features, but not about some new expert system. Keyword search engine enriched by expert system features is aimed at using data from the Web and rules, which are generated from the web resources such as OWL ontologies and raw text of web sites.

**Acknowledgements.** This work has been supported by my family.

## References

1. GOOGLE. www.google.com
2. BING. www.bing.com
3. BAIDU. www.baidu.com
4. Verhodubs O.: Mutual transformation of information and knowledge (2016)
5. Jackson P.: Introduction to Expert Systems (3 ed.) (1998)
6. Т. А. Гаврилова, В. Ф. Хорошевский: *Базы знаний интеллектуальных систем. Учебник.* Санкт-Петербург: ПИТЕР (2000)
7. Expert systems. http://users.cs.cf.ac.uk/Dave.Marshall/AII/mycin.html
8. Web Ontology Language. https://www.w3.org/TR/owl-features/
9. Verhodubs O., Grundspenkis J.: Evolution of ontology potential for generation of rules (2012). doi: 10.1145/2254129.2254201
10. Verhodubs O.: Ontology as a source for rule generation (2014)
11. Metarule. https://www.thefreedictionary.com/metarule
12. Conflict resolution. http://www.jhigh.co.uk/Higher/expert_systems/conflict_resolution.html
13. Verhodubs O., Grundspenkis J., Towards the Semantic Web Expert System (2011)
14. Apache Jena. https://jena.apache.org/
15. Verhodubs O.: Adaptation of the Jena framework for fuzzy reasoning in the Semantic Web Expert System (2014)